\documentstyle[times,epsfig]{jaa}
\newcommand{\etal}{{ et al. }} 

\newcommand{\be}{\begin{equation}}
\newcommand{\ee}{\end{equation}}

\begin{document}
\title[Black hole masses of blazars]{Estimating black hole masses of blazars} 
\author[X.-B. Wu]%
       {Xue-Bing Wu\thanks{e-mail:wuxb@pku.edu.cn}$^1$, F. K. Liu$^1$,  M. Z. Kong$^{2}$, R. Wang$^3$,
J. L. Han$^4$ \\ 
        $^1$Department of Astronomy, Peking University, Beijing 100871, China\\
$^2$ Department of Physics, Hebei Normal University, Shijiazhang 050016, China\\
$^3$ Steward Observatory, University of Arizona, Tucson, AZ 85721, USA\\
$^4$ National Astronomical Observatories, Chinese Academy of Sciences, 
Beijing 100012, China\\
}
\maketitle
\label{firstpage}
\begin{abstract}
Estimating black hole masses of blazars is still a big challenge.
Because of the contamination of jets, using the previously suggested 
size -- continuum luminosity relation can overestimate the broad line
region (BLR) size and black hole mass for radio-loud AGNs, including blazars. 
We propose a new 
relation between the BLR size and $H_{\beta}$ emission line luminosity and
present evidences for using it to get more accurate black hole 
masses of radio-loud AGNs. For extremely radio-loud AGNs such as blazars with
weak/absent  emission lines, we suggest to use the fundamental plane
relation of their elliptical host galaxies to estimate the central velocity
dispersions and black hole masses, if their velocity
dispersions are not known but the host galaxies can be mapped. The
black hole masses of some well-known blazars, such as OJ 287, AO 0235+164 and 
3C 66B, are obtained using these two methods and the M - $\sigma$ relation. 
The implications of their black hole masses on other
related studies are also discussed.
\end{abstract}

\section{Introduction}
\label{sec:intro}
In the last two decades, dynamical measurements have clearly revealed that 
supermassive black holes exist in the center of nearby galaxies. 
 However, dynamical methods 
can not be applied to most of AGNs because they are too bright.
Currently the most reliable method for AGN black hole mass estimation
is the reverberation mapping. Using this technique, the BLR size
can be measured using the time lag between the
variations of continuum and emission line fluxes. The black hole mass can be
derived from the BLR size and the characteristic velocity 
 using the virial law. So far, reverberation mapping
studies have yielded black hole masses of about 40
Seyfert 1 galaxies and nearby  quasars (
Kaspi \etal 2000, 2005; Peterson \etal 2004). 
 Using the observed data of these  reverberation 
mapping AGNs, an empirical relation between the BLR size ($R$) and  continuum 
luminosity at 5100$\AA$ ($L_{5100\AA}$) has been derived by Kaspi et al. 
(2000, 2005), which has been frequently adopted to estimate the BLR size and 
the black hole masses for large samples of AGNs, including radio-loud 
quasars. 
However, the optical continuum luminosity of  radio-loud AGNs may not
be a good indicator of ionizing luminosity. 
Powerful jets of blazar-type AGNs may significantly contribute to the 
optical continuum luminosity.  
Therefore, using the  $R-L_{5100\AA}$ relation 
 may significantly overestimate the actual
BLR size and the black hole mass of radio-loud AGNs. 
In addition, another tight 
correlation between black hole mass and bulge velocity
dispersion ($\sigma$) has been found for nearby galaxies 
(Tremaine et al. 2002) and for a few Seyfert galaxies as well (Ferrarese et al.
 2001).   Such a  relation suggests a
possibility to estimate the black hole masses of AGNs using the
measured bulge velocity dispersions.  Especially for BL Lacertae objects,
the reverberation mapping technique cannot be applied  because they 
show no or weak emission lines in their optical spectra. Using
the  M$_{BH}$-$\sigma$ relation may be the only way to derive
their black hole masses, though measuring $\sigma$ is  
possible only for nearby sources. We have to seek for other methods for most
blazars because
the velocity dispersions of their host galaxies are too difficult 
to be measured.

In this paper we report our progress on estimating the 
black hole masses of radio-loud AGNs using a new BLR size -- $H_{\beta}$ 
emission line luminosity relation and the fundamental plane relation of
the elliptical host galaxies. The results on some well-known blazars are
also presented.

\section{The  BLR size -- H$_\beta$ luminosity relation}
Using the available data of BLR sizes 
and H$_\beta$ fluxes for 34 AGNs in the reverberation mapping studies 
(Kaspi et al. 2000), we  
investigated the relation between the BLR size and the
H$_\beta$ emission line luminosity (Wu \etal 2004). An empirical
relation between the BLR size 
and H$_\beta$ luminosity was derived as:
\be
\rm{Log}~R~(\rm{lt-days}) = (1.381\pm0.080)+(0.684\pm0.106) Log~(L_{H_\beta}/10^{42}~ergs~s^{-1}) .
\ee
The Spearman's rank correlation coefficient of this relation is 0.91. 
In Fig. 1 we show the dependence of the BLR size 
on $L_{H_\beta}$ and $L_{5100\AA}$. 
Obviously these two relations are similar, which means that the 
$R-L_{H_{\beta}}$ relation can be an alternative of the  $R-L_{5100\AA}$ 
relation in estimating the BLR size for radio-quiet AGNs. We applied both
the $R-L_{H_{\beta}}$ and  $R-L_{5100\AA}$ relations to estimate
the black hole masses of 87 radio-loud quasars and compare them in  
Fig. 2. Evidently the masses obtained with
the  $R-L_{H_\beta}$ relation are systematically lower that those obtained with
the $R-L_{5100\AA}$ relation for some extremely radio-loud quasars. 
The difference between two black hole mass estimates is smaller when the 
radio-loudness is small but becomes
larger as the radio-loudness increases. For some  individual quasars with 
higher radio-loudness, the black hole mass estimated with the  $R-L_{5100\AA}$ 
relation can be 3$\sim$10 times larger than that estimated with the 
$R-L_{H_\beta}$ relation.   
Recently Kong \etal (2006) also extended such a study to the broad UV emission
lines MgII and CIV, and obtained the BLR size -- UV emission line luminosity 
relations. Our results demonstrated that using the 
BLR size -- emission line luminosity 
relations  can avoid the overestimations of the black hole masses for blazar-like radio-loud
AGNs.
\begin{figure}[!ht]
\psfig{file=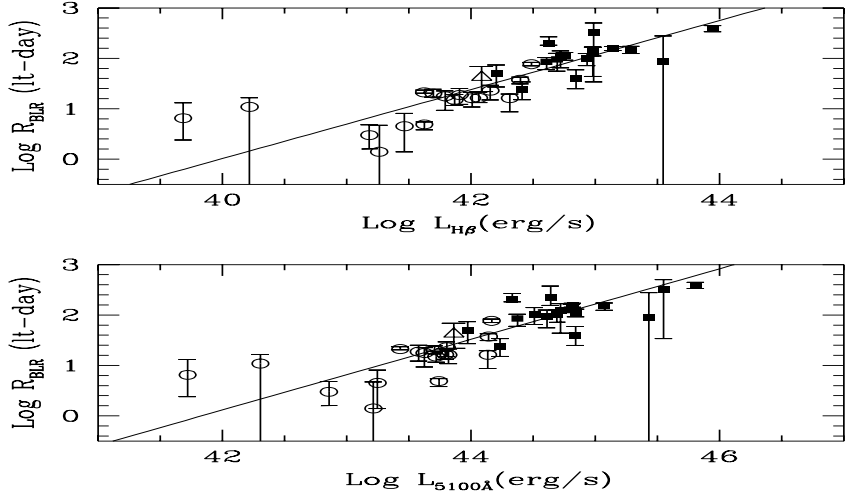,width=13cm,height=7cm,angle=0}
\caption{The $R-L_{H_\beta}$ relation and the $R-L_{5100\AA}$ 
relation. The open and filled symbols denote Seyferts and
quasars respectively. The figure is taken from Wu \etal (2004).}
\end{figure}

\begin{figure}[!ht]
\psfig{file=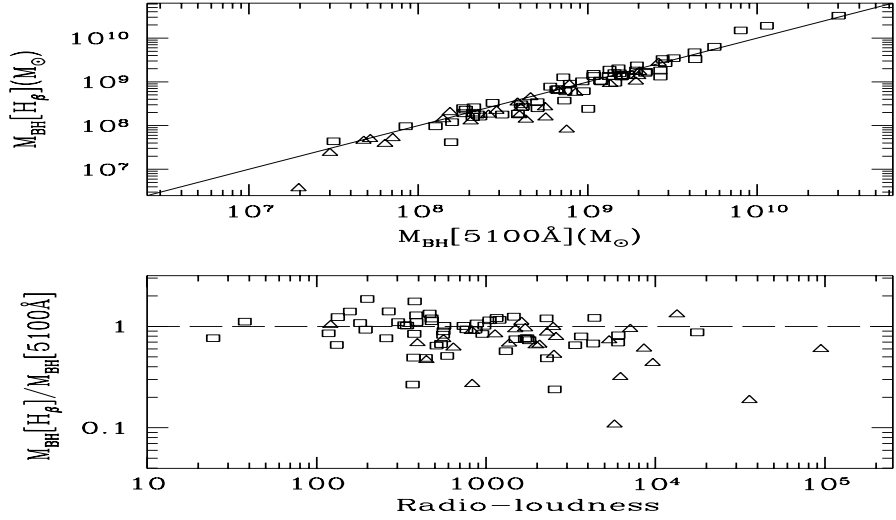,width=13cm,height=7cm,angle=0}
\caption{Comparison of the black hole masses of radio-loud quasars 
estimated with two R--L  relations, and the dependence of the black hole mass 
difference
on radio loudness. The figure is taken from Wu \etal (2004).}
\end{figure}

\section{Black hole masses estimated from the fundamental plane relation
of AGN elliptical host galaxies}
For radio-loud AGNs such as BL Lacs with
weak/absent emission lines, the emission line based methods for black hole
mass estimations can not apply. Directly measuring their
stellar velocity dispersion is also difficult. However, 
the host galaxies of BL Lacs are virtually ellipticals. 
 It is well known for  
ellipticals that three observables, namely the effective radius ($R_e$), 
the average surface
brightness ($<\mu_e>_R$ in R-band) and the central velocity dispersion 
($\sigma$), follow a tight fundamental plane relation. 
For about 300 normal ellipticals and radio galaxies,
Bettoni et al. (2001) found that the fundamental plane can be robustly 
described as 
\begin{equation}
\log R_e = (1.27\pm0.04) \log \sigma + (0.326\pm0.007) <\mu_e>_R
- 8.56\pm0.06,
\end{equation}
This  relation provides us
another way to estimate the central velocity dispersions and then
the black hole masses of AGNs (Wu, Liu \& Zhang 2002).

\begin{figure}[!ht]
\psfig{file=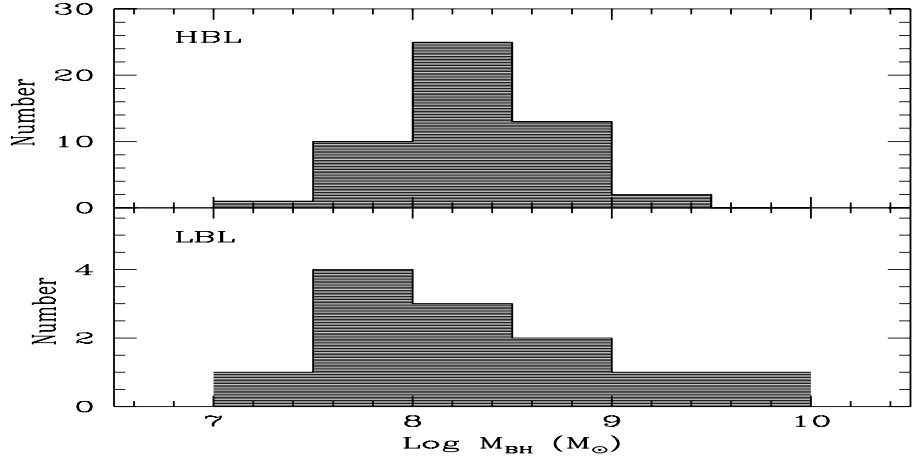,width=12cm,height=6cm,angle=0}
\caption{Histograms of the derived black hole mass distribution 
of HBLs and LBLs. The figure is taken from Wu \etal (2002).}
\end{figure}

\begin{figure}[!ht]
\psfig{file=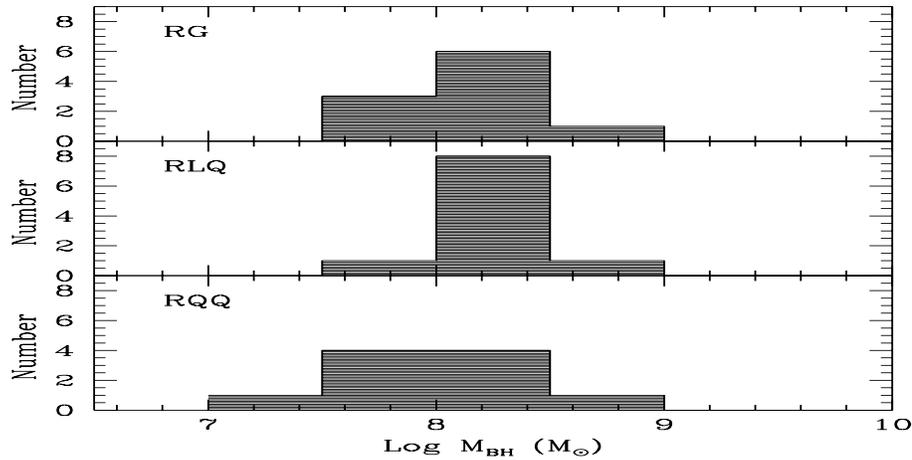,width=12cm,height=6cm,angle=0}
\caption{Histograms of the derived black hole mass distribution 
of radio galaxies, 
radio-loud and radio-quiet quasars. The figure is taken from Wu \etal (2002).}
\end{figure}

Using the imaging data of  BL Lacs obtained from the HST snapshot
survey (Urry \etal 2000), we adopted the fundamental plane relation to 
estimate the central velocity dispersions and  black hole masses of
51 high-frequency peaked BL Lacs (HBLs) and 12 low-frequency peaked BL Lacs
(LBLs). Our results show no significant difference in
the black hole masses between HBLs and LBLs (see  
Fig. 3). We also applied the same 
method  to 10 radio galaxies (RGs), 10 radio-loud
quasars (RLQs) and 13 radio-quiet quasars (RQQs) which have been imaged by
HST (McLure et al. 1999), we found that there are no significant differences 
in the 
black hole masses among these different types of AGNs with elliptical host
galaxies (see Fig. 4).  

\section{Black hole masses of three well-known blazars}
\subsection{AO 0235+164}
Liu, Zhao, \& Wu (2006) 
applied the $R-L_{H_\beta}$ relation suggested by Wu et al. (2004) to a well studied BL Lac object AO 
0235+164 and estimated its black hole mass as $5.8 \times 10^8 M_\odot$, which
is consistent with the mass ($3.6 \times 10^8 M_\odot$) estimated from 
the $M_{BH}-\sigma$ relation by 
taking the narrow emission line width as a surrogate of $\sigma$, but is much 
smaller than the mass ($1.5 \times 10^9 M_\odot$) obtained from the 
$R-L_{5100\AA}$ relation. This again demonstrates that using the previous
$R-L_{5100\AA}$ relation can overestimate the black hole masses of blazar-like
AGNs. Our study indicates that the black hole mass of AO 0235+164 is most likely
around $5 \times 10^8 M_\odot$.

\subsection{OJ 287}
We applied the fundamental plane relation of ellipticals to a well-known BL Lac object OJ 287, which is a possible supermassive black hole binary system. Liu \& Wu (2002) estimated its primary black hole mass 
to be about
$4\times10^8M_\odot$, which is consistent with the upper limit ($10^9M_\odot$)
obtained by Valtaoja
et al. (2000) based on a binary
black hole model for OJ 287. This value is also 
within the range of typical black hole masses for BL lac objects (Wu, Liu \&
Zhang 2002). We noticed that our derived black hole mass is
much smaller than $1.8\times10^{10}M_\odot$, which is required in a new
precessing binary black hole model of OJ 287 (Valtonen 2007).  

\subsection{3C 66B}
3C~66B is a well-known nearby FR I radio galaxy. The central velocity dispersion of 3C~66B has been obtained by Balcells et al. 
(1995) as  348$\pm$29 km/s based on their spectroscopic observations. Using the M - $\sigma$ relation
suggested by Tremaine et al. (2002), we estimate that $\log(M/M_\odot)
=9.097\pm0.175$, namely the mass of the 
central black hole of 3C~66B is about $1.25\times 10^9 M_\odot$. This is 
within the range of typical black hole masses for radio galaxies (Wu, Liu \&
Zhang 2002), but one-order of magnitude smaller than the 
maximum mass estimated
by Sudou et al. (2003), $5 \times 10^{10} M_\odot$,  which was obtained by
modeling the observed orbital motion with a binary black hole model. 
A very recent study by Iguchi et al. (2010) supported that the larger black hole mass is $(1.2^{+0.5}_{-0.2}) \times 10^9$ $M_{\odot}$, which is consistent
with the value obtained by the  M - $\sigma$ relation.
Janet et al. (2004) adopted the black hole mass of 3C~66B obtained by Sudou et al. (2003) to calculate the gravitational wave emission from 3C~66B and argued that  the binary black hole model can be excluded because observations from the
Pulsar Timing Array (PTA) did not detect the gravitational wave from 3C~66B. 
However, if the smaller black hole mass is adpted, the gravitational wave emitted from 3C~66B will be weaker than the current PTA detection limit and the binary
black hole model can not be excluded.

\section{Summary}
We proposed to use the BLR size -- emission line luminosity relation and 
the fundamental
plane relation of the elliptical host galaxies to estimate the
black hole masses of radio-loud
AGNs. We demonstrated that with the first relation we can get more
accurate black hole mass estimates for radio-loud AGNs than using the usual
$R-L_{5100\AA}$ relation, and for some radio-loud AGNs such as BL Lacs
the second method is probably the only available one for their
black hole mass estimations in the case when  directly
measuring the stellar velocity dispersions is difficult. We have adopted these
methods, as well as the M - $\sigma$ relation, to estimate the black hole masses of three well-known blazars, AO 0235+164, OJ 297 and 3C 66B, and the results are helpful to other related study about their physics nature.  Finally, we would
like to mention that these two methods can be also applied to estimate the
black hole masses of high redshift AGNs, including blazars, with high quality spectroscopic
and imaging observations.   

This work is supported
by the NSFC Grants  (No. 10473001, 10573001,
 10525313 and 11033001) and the 973 program (No. 2007CB815405).



\begin{thebibliography}{10}
\bibitem{B95} Balcells, M. et al. 1995, A\&A, 302, 665
\bibitem{B01} Bettoni, D.  et al. 2001, A\&A, 380, 471
\bibitem{F01} Ferrarese, L. \etal 2001, ApJ, 555, L79
\bibitem{I10} Iguchi, S., Takeshi, O., \& Sudou, H. 2010, ApJ, 724, L166
\bibitem{J04} Jenet, F.A., \etal 2004, ApJ, 606, 799
\bibitem{K00} Kaspi, S. et al. 2000, ApJ, 533, 631
\bibitem{K05} Kaspi, S. et al. 2005, ApJ, 629, 61 
\bibitem{K06} Kong, M. Z., Wu, X.-B., Wang, R., Han, J. L. 2006, Chin. J. 
Astron. Astrophys. 6, 396 
\bibitem{LW02} Liu, F. K., \& Wu, X.-B., 2002, A\&A, 388, L48
\bibitem{LZW06}Liu, F. K., Zhao, G., \& Wu, X.-B., 2006, ApJ, 650, 749
\bibitem{R14} McLure, R. J., et al. 1999, MNRAS, 308, 377
\bibitem{P04} Peterson, B. M.  \etal 2004, ApJ, 613, 682
\bibitem{S03} Sudou, H. et al., \etal 2003, Science, 300, 1263
\bibitem{T02} Tremaine, S. et al. 2002, ApJ, 574, 740 
\bibitem{U00} Urry, C. M. \etal 2000, ApJ, 532, 816
\bibitem{V00} Valtaoja, E. \etal. 2000, ApJ, 531, 744
\bibitem{V07} Valtonen, M.J., 2007, ApJ, 659, 1074   
\bibitem{wulz}Wu, X.-B., Liu, F. K., Zhang, T. Z. 2002, A\&A, 389, 742 
\bibitem{wu04}Wu, X.-B., Wang, R., Kong, M. Z., Liu, F. K., Han, J. L., 2004, A\&A, 424, 793


\end{thebibliography}
\end{document}